\documentclass[letterpaper,prb,twocolumn]{revtex4}
\usepackage{graphicx,bm}
\usepackage{amsmath,amssymb}

\begin{document}

\title{Zero-mode anomaly in the RKKY interaction on bipartite lattices}
\author{Tsung-Cheng Lu and Hsiu-Hau Lin}
\affiliation{
Department of Physics, National Tsing Hua University, Hsinchu 30013, Taiwan
}
\date{October 4, 2014}

\begin{abstract}
Carrier-mediated Ruderman-Kittel-Kasuya-Yosida (RKKY) interaction plays an important role in itinerant magnetism. There have been intense interest on its general trend on bipartite lattice with particle-hole symmetry. In particular, recently fabricated graphene is well described by the honeycomb lattice within tight-binding approximation. We use SUSY quantum mechanics to study the RKKY interaction on bipartite lattices. The SUSY structure naturally differentiate the zero modes and those paired states at finite energies. The significant role of zero modes is largely ignored in previous literature because their measure is often zero in the thermodynamic limit. Employing both real-time and imaginary-time formalism, we arrive at the same conclusion: The RKKY interaction for impurity spins on different sublattices is always antiferromagnetic. However, for impurity spins on the same sublattice, the carrier-mediated RKKY interaction is not always ferromagnetic. Only in the absence of zero modes, the sign rule on the bipartite lattice holds true. Our finding highlight the importance of the zero modes in bipartite lattices. Their significance needs further investigation and may lead to important advances in carrier-mediated magnetism.
\end{abstract}

\maketitle

\section{Introduction}
Carrier-mediated exchange interaction between two impurity spins, know as Ruderman-Kittel-Kasuya-Yosida (RKKY) interaction, plays a fundamental role in itinerant magnetism and has many critical applications in spintronics. Because the RKKY interaction is dictated by the Fermi surface, the particle-hole symmetry in bipartite lattices will mark peculiar features in the spatial profile of the mediated interactions. 

In a recent paper, Saremi\cite{Saremi2007} showed that the RKKY interaction on a bipartite lattice is always ferromagnetic on the same sublattice while antiferromagnetic on opposite sublattices. This theorem is widely applied to graphene\cite{Novoselov2004,Geim2007,Neto2009,Yazyev2010} because its band structure is well approximated the honeycomb lattice (a bipartite one) within the tight-binding approximation. In addition, the relativistic dispersion in graphene also changes the power-law exponent in the long-distance limit.

Neglecting the spin-orbit interaction momentarily, it is generally accepted\cite{Black-Schaffer2010,Sherafati2011,Kogan2011,Power2013} that the itinerant carriers mediate the ferromagnetic exchange coupling between two impurity spins on the same sublattice. On the other hand, if the impurity spins sit on different sublattices, the mediated exchange coupling turns antiferromagnetic. However, Bunder and Lin\cite{Bunder2009} later pointed out that the ``sign rule" on the bipartite lattice is ruined when the zero modes are present. There are debates on the discrepancy, potentially arising from different theoretical approaches. In addition, these zero modes are closely related to recently found magnetism in graphene and related materials.\cite{Lieb1989,Yazyev08,Brey07,Wunsch08,Yang08,Dutta08,Jung09,Potasz12}

Inspired by the current debates and confusions in the community, we investigate the carrier-mediated RKKY interaction on bipartite lattices again, with emphasis on treating the zero modes properly. It turns out that the supersymmetric (SUSY) quantum mechanics is the optimal approach to make the symmetry for the finite-energy modes explicit. Meanwhile, the zero modes arise from the null space of the supercharge and give anomalous contribution to the RKKY interaction. Because the zero modes are annihilated by the supercharge operator, the corresponding wave functions only show up on one sublattice and completely vanish on the other sublattice -- the so-called nodal structure. The presence of these zero modes destroy the sign rule proven by Saremi before. Why does the sign rule work so well in graphene then? The reason turns out to be relatively simple: the measure of these zero modes in graphene goes to zero in the thermodynamic limit and the anomaly becomes invisible.

The rest of the paper is organized in the following way. In Section II, we introduce the SUSY approach and describe the difference between zero modes and those paired states at finite energies. In Section III, the carrier-mediated RKKY interaction on the bipartite lattice is derived by the real-time formalism. All major results of the paper are derived in this section. In Section IV, a brief guide for derivations via imaginary-time approach is presented. Both real-time and imaginary-time formalism gives the same results. We also pointed out an important mistake in the literature. At the end, the conclusion is presented.

\section{bipartite lattice in SUSY form}

We start with the simple hopping Hamiltonian on a bipartite lattice,
\begin{equation}
\label{eq:hamiltonian}
H_=\displaystyle\sum_{i,j}t_{ij}c^\dagger_ic_j,
\end{equation}
where $c_i, c^\dag_i$ are annihilation and creation operators on lattice site $i$ and the hopping amplitude $t_{ij} = 0$ when $i, j$ belong to the same sublattice. For simplicity, we assume these hopping amplitudes are real. The hermiticity of the Hamiltonian requires the hopping matrix to be symmetric $t_{ij} = t_{ji}$. It is insightful to rewrite the hopping Hamiltonia into SUSY form. Suppose there are $N_A$ sites for sublattice $A$ and $N_B$ sites for sublattice $B$. By rearranging lattice indices, the Hamiltonian can be cast into the standard SUSY form,
\begin{equation}
\label{eq:susy}
H=\begin{pmatrix}
  0 & Q \\
  Q^{\dag} & 0
  \end{pmatrix},
\end{equation}
where $Q$ is the supercharge operator of dimensions $N_A \times N_B$. In general, $N_A$ and $N_B$ are not necessarily equal and the zero modes live in the corresponding null space of $Q$ and/or $Q^\dag$.

Due to its quadratic nature, it is straightforward to work out all finite-energy modes. The SUSY structure of the hopping Hamiltonian ensures these states appear in pair with the simple sign rule,
\begin{equation}
\label{eq:ph symm wave fn}
\phi_{\bar{n}}(i)=\epsilon_i \phi_n (i).
\end{equation}
Here $\phi_{n}, \phi_{\bar{n}}$ are the wave functions of the finite-energy states. Note that $\epsilon_{i} = \pm 1$ for sublattice $A$ and $B$ respectively and $n, \bar{n}$ represent the paired quantum number with opposite energies. But, the zero modes do not appear in pairs.  Besides, because they are annihilated by either $Q$ or $Q^\dag$, their wave functions only show up on one sublattice. The direct consequence is that the sign rule for the Green's function breaks down,
\begin{equation}
\label{eq:ph symm gre fn}
\langle c_ic^{\dagger}_j \rangle \neq
\epsilon_i \epsilon_j \langle c_i^{\dagger}c_j \rangle.
\end{equation}
The above relation only holds when the anomaly due to zero modes vanishes. Previous studies made use of this relation, leading to the incorrect conclusion for the RKKY interaction on bipartite lattices.

Let us try to elaborate on the details. If there are no zero modes, all states appear in pairs. Expanding the lattice operator in eigenbasis, $c_i=\sum_m \phi_m(i)c_m$, the Green's function can be expressed as
\begin{equation}
\begin{aligned}
\langle c_ic^{\dagger}_j \rangle&
=\sum_{m,n} \phi_m(i)\phi_n(j) \langle c_mc^{\dagger}_n \rangle,
\\
&=\sum_{m,n} \phi_{\bar{m}}(i)\phi_{\bar{n}}(j) \langle c_{\bar{m}}c^{\dagger}_{\bar{n}} \rangle.
\end{aligned}
\end{equation}
In the second line, the dummy indices are changed to the opposite paired quantum numbers. Making use of the particle-hole symmetry,
\begin{equation}
\begin{aligned}
\langle c_ic^{\dagger}_j \rangle & =
\sum_{m,n} \epsilon_i \epsilon_j\phi_{m}(i)\phi_{n}(j) \langle c_m^{\dagger} c_n \rangle
=\epsilon_i \epsilon_j \langle c_i^{\dagger}c_j \rangle.
\end{aligned}
\end{equation}
The above relation holds true when the zero modes are absent. In the presence of zero modes, the above derivation fails and the sign rule for the RKKY interaction is no longer valid.

\section{real-time formalism}

There have been debates over the sign rule of the RKKY interaction on a bipartite lattice. Some studies attribute the discrepancy to the employments of the real-time or the imaginary-time formalism. Thought both approaches are fundamental and can be found in many standard textbooks, it is insightful to walk through the key steps. The carrier-mediated RKKY interaction between two impurity spins $\bm{S}_i$ and $\bm{S}_{j}$ is
\begin{equation}
H_{\rm RKKY}=J_{ij} \bm{S}_i \cdot \bm{S}_j.
\end{equation}
The mediated exchange coupling is directly related to the spin susceptibility within the linear response theory,
\begin{equation}
J_{ij}=-J^2\chi_{ij}^R(\omega=0),
\end{equation}
where $J$ is the direct exchange coupling between the itinerant spin densities and the impurity spins. Furthermore, retardation effects are ignored so that only the static susceptibility is involved here,
\begin{equation}
\label{eq:staticchi}
\chi^R_{ij}(\omega = 0)=\int^{\infty}_{0}\!dt\: \frac{i}{2} \left< \left[ s^-_i(t),s^+_j(0) \right]\right> e^{-\eta t}.
\end{equation}
The spin correlation function can be decomposed into product of single-particle Green's function,
\begin{equation}
\begin{aligned}
\left<s^-_i(t)s^+_j(0)\right>&=\left< c^{\dagger}_i(t)c_j(0) \right>\left< c_i(t)c^{\dagger}_j(0) \right>\\
& \hspace{-2cm}=\displaystyle\sum_{m,n}W_{mn}(i,j)n_F(\xi_m)(1-n_F(\xi_n))e^{i(\xi_m-\xi_n)t},
\end{aligned}
\end{equation}
where $W_{mn}(i,j)=\phi_m(i)\phi_m(j)\phi_n(i)\phi_n(j)$. After some algebra, the retarded susceptibility can be expressed as
\begin{equation}
\begin{aligned}
\label{eq:realchi}
\chi_{ij}^R &=-\displaystyle\sum_{m,n}W_{mn}(i,j)\:
\frac{n_F(\xi_m)[1-n_F(\xi_n)](\xi_m-\xi_n)}{(\xi_m-\xi_n)^2+\eta ^2}\\
&=-\displaystyle\sum_{\xi_m\neq\xi_n}W_{mn}(i,j)\frac{n_F(\xi_m)[1-n_F(\xi_n)]}{\xi_m-\xi_n}.
\end{aligned}
\end{equation}
For simplicity, let us focus on the zero temperature first. Making use of the SUSY structure, the above summation can be separated into the normal and anomalous parts: $\chi^R_{ij} = \chi^n_{ij} + \chi^a_{ij}$. The normal part does not involve any zero modes,
\begin{equation}
\begin{aligned}
\label{eq:nonzeromode}
\chi_{ij}^n
&=\epsilon_i\epsilon_j\displaystyle\sum_{\xi_m,\xi_n>0}
\int^{\infty}_0 \!dt\:  W_{mn}(i,j) e^{-(\xi_m+\xi_n)t}\\
&=\epsilon_i\epsilon_j\int^{\infty}_0 dt \left (\displaystyle\sum_{\xi_m>0}\phi_m(i)\phi_m(j) e^{-\xi_m t}\right)^2.
\end{aligned}
\end{equation}
Because the integrand is positive definite, the sign of $\chi^n$ is solely determined by the factor $\epsilon_i \epsilon_j$. For the same sublattice, $\epsilon_i \epsilon_j=1$ and the mediated exchange coupling is ferromagnetic. For opposite sublattices, $\epsilon_i \epsilon_j = -1$ and the mediated coupling is antiferromagnetic.

However, the presence of zero modes give rise to anomalous contributions,
\begin{equation}
\begin{aligned}
\label{eq:zeromodecontribution}
\chi_{ij}^a=&
\sum_{\xi_m=0,\xi_n>0} \frac12 \frac{W_{nm}(i,j)}{\xi_n-\xi_m}
+\sum_{\xi_m=0,\xi_n<0}\frac{1}{2}\frac{W_{mn}(i,j)}{\xi_m-\xi_n}
\\
&=\frac{1+\epsilon_i \epsilon_j}{2} \sum_{\xi_m=0,\xi_n>0} W_{mn}(i,j)\: \xi_n^{-1}.
\end{aligned}
\end{equation}
The sign rule for the anomalous part $\chi^a_{ij}$ is different from the normal one $\chi^n_{ij}$. First of all, the anomalous contribution vanishes if $\epsilon_i \epsilon_j=-1$. That is to say, the retarded susceptibility $\chi_{ij}^{R}$ always gives rise to antiferromagnetic coupling between two impurity spins on different sublattices. But, for impurity spins sitting on the same sublattice, the normal part $\chi^n > 0$ is positive yet the anomalous part $\chi^a$ depends on microscopic details. In consequence, the RKKY interaction is not necessarily ferromagnetic and the sign rule breaks down.

As a demonstrating example, we consider one dimensional chain with $N$ sites and uniform hopping $t$. The corresponding Harper equation is
\begin{equation}
-t\phi(x-1) - t\phi(x+1)=E\phi(x),
\end{equation}
where the site index run through the bulk values $x=2,3,\cdots,N-1$. The Harper equations for $x=1, N$ are different due to open boundary conditions and can be solved by the fictitious fields $\phi(0)=\phi(N+1)=0$. Analytic solutions for the energy dispersion and the corresponding wave functions are
\begin{equation}
\begin{aligned}
\label{eq:energyspectrum}
\xi_m= - 2t \cos \left(\frac{m\pi}{N+1} \right),
\\
\phi_m(x)=A\: \sin\left(\frac{m\pi}{N+1}x \right),
\end{aligned}
\end{equation}
where $m=0,1,2,...N$, and $A$ is the normalization constant. The retarded spin susceptibility $\chi^R_{ij}$ can be computed straightforwardly as shown in Fig. 1. Because the RKKY interaction $J_{ij} = -J^2 \chi^R_{ij}$, it is clear that the carrier-mediated exchange coupling is \textit{always} antiferromagnetic when two impurity spins are on different sublattices. On the other hand, the RKKY interaction for impurity spins on the same sublattice does not obey any concrete sign rule: it is ferromagnetic at short distances and gradually turns antiferromagnetic in the long distance limit.

\begin{figure}
\centering
\includegraphics[width=7.5cm]{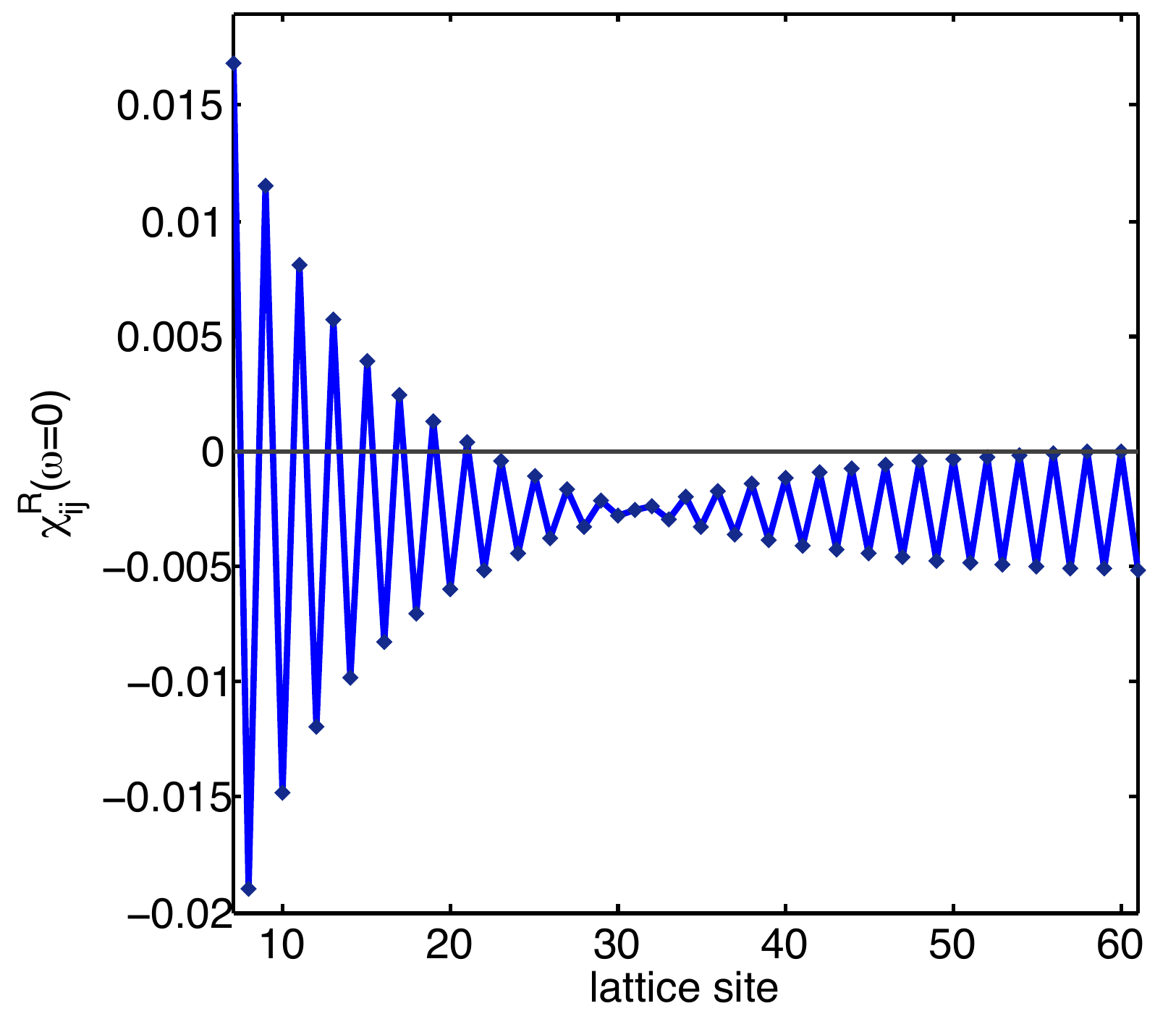}
\caption{The retarded spin susceptibility $\chi^R_{ij}(\omega=0)$ for the one dimensional chain with one impurity spin placed at the boundary site $x=1$. The usual oscillatory trend remains even for discrete lattice structure. The RKKY interaction for opposite sublattices follows the sign rule and persists to be antiferromagnetic. But, there is no such sign rule when both impurity spins are on the same site.}\label{AB}
\end{figure}

\section{Imaginary-time formalism revised}
The main results of the paper has been obtained by the real-time formalism already. However, it is helpful to compare with the imaginary-time formalism as well. The Matsubara Green's function for the spin susceptibility is
\begin{equation}
\label{eq:imaginarychi}
\chi_{ij}(i\Omega_n)=\int_0^{\beta} \!d\tau\:
e^{i\Omega_n\tau}\frac{1}{2}\left< T_{\tau}s_i^-(\tau)s_j^+(0) \right>,
\end{equation}
where $T_{\tau}$ is the time-ordering operator in imaginary time. If the integration over the imaginary time is carried out first, we end up with the standard Lehmann decomposition. It is well known that the analytic continuation $i \Omega_n \to \omega + i\eta$ of the Lehmann series connects the Matsubara Green's function to the retarded one in real time. The integration is straightforward and it indeed leads to the same results as derived from the real-time formalism.

The mistake arises if the analytic continuation is imposed first. The spin susceptibility in the imaginary-time formalism now takes a rather simple-looking form, 
\begin{equation}
\begin{aligned}
\label{eq:wrongans}
\chi_{ij}=\int_0^{\beta}d\tau\frac{1}{2} \langle s_i^-(\tau)s_j^+(0) \rangle.
\end{aligned}
\end{equation}
The time-ordering operator can be dropped because the imaginary time runs between 0 and $\beta$. The spin correlation function can be expressed in the eigenbasis,
\begin{equation}
\begin{aligned}
\label{eq:imaginary correlation}
\left<s^-_i(\tau)s^+_j(0)\right>&=
\left< c^{\dagger}_i(\tau)c_j(0) \right>\left< c_i(t)c^{\dagger}_j(0) \right>
\\
& \hspace{-2cm}
=\displaystyle\sum_{m,n}W_{mn}(i,j)n_F(\xi_m)[1-n_F(\xi_n)]e^{(\xi_m-\xi_n)\tau}.
\end{aligned}
\end{equation}
The above sum can be sorted into $\xi_m=\xi_n$ and $\xi_m\neq\xi_n$ parts. The first part with equal energy gives
\begin{equation}
\label{eq:}
\chi_{ij}^0=\frac{\beta}{2}\displaystyle\sum_{\xi_m=\xi_n}
W_{mn}(i,j)n_F(\xi_m)[1-n_F(\xi_n)].
\end{equation}
After some algebra, the second part turns out to be the retarded spin susceptibility. That is to say,
\begin{equation}
\chi_{ij} = \chi^R_{ij} + \chi_{ij}^0.
\end{equation}
The discrepancy between $\chi_{ij}$ and $\chi_{ij}^R(\omega=0)$ always exists at all temperatures. The sign rule is strictly correct for $\chi_{ij}$ but it is not the realistic spin susceptibility measured in experiments. It is worth emphasizing that there is nothing wrong with the imaginary-time formalism as long as the analytic continuation is correctly performed.

\section{Conclusion}
We use SUSY quantum mechanics to study the RKKY interaction on bipartite lattices. The SUSY structure naturally differentiate the zero modes and those paired states at finite energies. Employing both real-time and imaginary-time formalism, we arrive at the same conclusion: The RKKY interaction for impurity spins on different sublattices is always antiferromagnetic. However, for impurity spins on the same sublattice, the carrier-mediated RKKY interaction is not always ferromagnetic. Only in the absence of zero modes, the sign rule on the bipartite lattice holds true. Our finding highlight the importance of the zero modes in bipartite lattices. Their significance needs further investigation and may lead to important advances in carrier-mediated magnetism.

We acknowledge supports from the Ministry of Science and Technology in Taiwan through grant MOST 103-2112-M-007-011-MY3. Financial supports and friendly environment provided by the National Center for Theoretical Sciences in Taiwan are also greatly appreciated.


\begin{thebibliography}{}

\bibitem{Saremi2007}
S. Saremi,
Phys. Rev. B \textbf{76}, 184430 (2007).


\bibitem{Novoselov2004}
K. S. Novoselov, A. K. Geim, S. V. Morozov, D. Jiang, Y. Zhang, S. V. Dubonos,
I. V. Grigorieva, A. A. Firsov,
Science, \textbf{306}, 666 (2004).

\bibitem{Geim2007}
A. K. Geim and K. S. Novoselov,
Nat. Mat. \textbf{6}, 183 (2007).

\bibitem{Neto2009}
A. H. Castro Neto, F. Guinea, N. M. R. Peres, K. S. Novoselov, and A. K. Geim,
Rev. Mod. Phys. \textbf{81}, 109 (2009).

\bibitem{Yazyev2010}
O. V. Yazyev,
Rep. Prog. Phys. \textbf{73}, 056501
(2010).


\bibitem{Black-Schaffer2010}
A. M. Black-Schaffer
Phys. Rev. B \textbf{81}, 205416 (2010).

\bibitem{Sherafati2011}
M. Sherafati and S. Satpathy,
Phys. Rev. B {\bf 83}, 165425 (2011).

\bibitem{Kogan2011}
E. Kogan,
Phys. Rev. B {\bf 84}, 115119 (2011).

\bibitem{Power2013}
S. R. Power and M. S. Ferreira,
Crystals {\bf 3}, 49 (2013).

\bibitem{Bunder2009}
J. E. Bunder and H.-H. Lin
Phys. Rev. B \textbf{80}, 153414 (2009).


\bibitem{Lieb1989}
E. H. Lieb,
Phys. Rev. Lett. {\bf 62}, 1201 (1989).

\bibitem{Yazyev08}
O. V. Yazyev, M. I. Katsnelson,
Phys. Rev. Lett. {\bf 100}, 047209 (2008).

\bibitem{Brey07}
L. Brey, H. A. Fertig, and S. Das Sarma
Phys. Rev. Lett. {\bf 99}, 116802 (2007).

\bibitem{Wunsch08}
B. Wunsch, T. Stauber, F. Sols, F. Guinea,
Phys. Rev. Lett. {\bf 101} 036803 (2008).

\bibitem{Yang08}
L. Yang, M.~L. Cohen and S.~G. Louie,
Phys. Rev. Lett. \textbf{101}, 186401 (2008).


\bibitem{Dutta08}
S. Dutta, S. Lakshmi, and S. K. Pati,
Phys. Rev. B {\bf 77}, 073412 (2008).

\bibitem{Jung09}
J. Jung and A.~H. MacDonald,
Phys. Rev. B {\bf 79}, 235433 (2009).

\bibitem{Potasz12}
P. Potasz, A. D. G\"u\c{c}l\"u, A. Wojs, and P. Hawrylak,
{Phys. Rev. B} {\bf 85}, 075431 (2012).



\end{thebibliography}
\end{document}